\documentclass[aps,twocolumn,prl,superscriptaddress]{revtex4-1}
\usepackage{xcolor} 
\usepackage{graphicx}
\usepackage{subfig}

\usepackage[normalem]{ulem}

\begin{document}
\title{Shifting computational boundaries for complex organic materials}

\author{R. Matthias Geilhufe}
\affiliation{Nordita,  KTH Royal Institute of Technology and Stockholm University, Roslagstullsbacken 23,  10691 Stockholm,  Sweden}
\email{matthias.geilhufe@su.se}
 
\author{Bart Olsthoorn}
\affiliation{Nordita,  KTH Royal Institute of Technology and Stockholm University, Roslagstullsbacken 23,  10691 Stockholm,  Sweden}

\author{Alexander V. Balatsky}
\email{balatsky@hotmail.com}
\affiliation{Nordita,  KTH Royal Institute of Technology and Stockholm University, Roslagstullsbacken 23,  10691 Stockholm,  Sweden}
\affiliation{Department of Physics, University of Connecticut, Storrs, CT 06269, USA}

\date{\today}

\maketitle

\textbf{Methodology adapted from data science sparked the field of materials informatics, and materials databases are at the heart of it. Applying artificial intelligence to these databases will allow the prediction of properties of complex organic crystals.}

Materials used today {often have complicated structure that makes their properties hard to predict, and so those with good specific functions are usually} discovered serendipitously. {And, while our computational capabilities are growing rapidly, there are still many situations that are too complex for accurate simulation.} As a result we have a gap between the {technological} demand for complex materials with desired properties and materials and properties that can be reliably calculated using modern ab initio methods. We focus on this tension in the context of organic materials, for two reasons. {Firstly,} carbon-based compounds exhibit an infinite configuration space and as a result an enormous potential for functional materials, and secondly due to the complexity of the compounds, the computational gap is particularly large for this class of materials. Organics have become a crucial part of modern technology, medicine, and chemistry, with wide-ranging applications, such as OLEDs, solar cells, or explosives, and were discussed in the context of molecular qubits, and spin-liquid physics \cite{gaita2019molecular,shimizu2003spin,layfield2014organometallic}.
\begin{table*}
    \centering
    \footnotesize
    \begin{tabular}{llp{5cm}l}
        \hline\hline
        Database & URL & Specifications & Ref. \\
        \hline
        Organic Materials Database & \url{https://omdb.mathub.io} &  organic molecular crystals, metal organic frameworks; band structure, DOS, magnetism, pattern matching, ML tools & \cite{borysov2017organic} \\
        Polymer Genome & \url{https://www.polymergenome.org/} &   experimental and computational data of polymers; ML tools & \cite{huan2016polymer}\\
        QM7, QM7b, QM8, QM9, ... & \url{http://quantum-machine.org/datasets/}  & small organic molecules; geometries minimal in energy, corresponding harmonic frequencies, dipole moments, polarizabilities, energies, enthalpies & \cite{ramakrishnan2014quantum}\\
        \hline\hline
    \end{tabular}
    
    \caption{A selection of ab initio databases and data sets dedicated for organic and metal organic materials.}
    \label{tab:databases}
\end{table*}
\begin{figure*}
    \centering
    \includegraphics{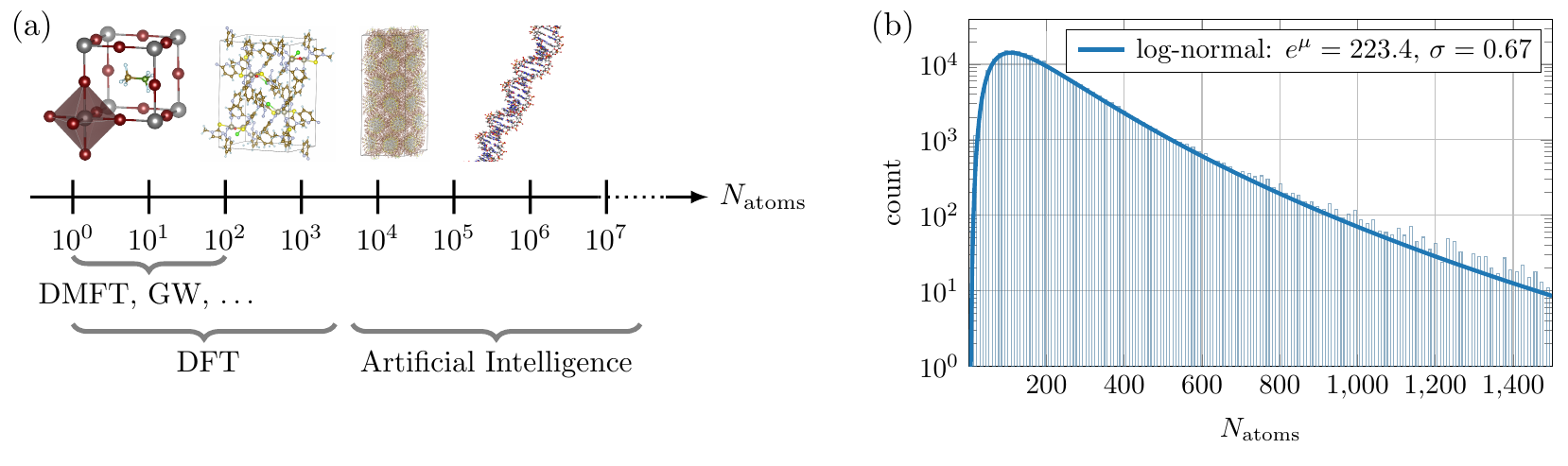}
    \caption{Organic molecular crystals and metal organic frameworks typically exhibit highly complex unit cells with hundreds of atoms present. Conventional ab intio approaches in particular taking into account strong correlation corrections are not feasible after a certain complexity is reached. Hence a new methodology is required. (a) outlines the range of current methodology and the number of atoms per unit cell for certain experimental crystal structures. (b) log-normal distribution of organic materials in the crystallographic open database.}
        \label{Fig:data_availability}
\end{figure*}
The field of materials informatics deals with the application of methodology adapted from computer science to materials research. With the exponential increase in computational power and storage capacity, numerous materials databases have emerged within the past decades, often  encouraged by focused  national or international strategic funding initiatives. Such databases {are} an exciting domain for data mining {and} machine learning. 

Within this trend we focus on recent computational databases for carbon-based materials (Table \ref{tab:databases}), for a broad overview of existing materials databases see, {for example,} Refs \cite{schleder2019dft,himanen2019data}. A prominent series of datasets storing calculated molecular properties is available {at} \url{http://quantum-machine.org/datasets/}. Calculated properties of crystalline organic materials are available via the Organic Materials Database (\url{omdb.mathub.io}). Properties of organic polymers can be found at the polymer genome (\url{https://www.polymergenome.org/}). {Subsequent work has shown that datasets from these resources are} valuable resources to train machine learning algorithms \cite{chmiela2018towards, olsthoorn2019band,doan2020machine}.

Data mining describes the quick exploration for materials with desired target properties within a materials database. This can be achieved by constraining the search space by specific parameters or {by using} pattern matching. One can use established heuristics  where a materials property is related to specific patterns in the available data such as the band structure.For example, topological semimetallicity {corresponds to the} occurrence of nodes in the band structure, or superconductivity {is linked to} peaks in the density of states at the Fermi level. Dedicated algorithms {that search} for similarities to specified initial patterns implemented into materials databases \cite{borysov2018online, geilhufe2018towards} provide a versatile tool to identify materials in almost no time, as has been shown for organic Dirac materials and topological materials \cite{geilhufe2017three, geilhufe2017data}, superconductors \cite{geilhufe2018towards}, energy materials \cite{zhang2019data}, magnetic materials \cite{hellsvik2020spin}, and quantum sensors \cite{geilhufe2018materials}. 

Given the great success of data-mining on materials databases for organics, a natural question arises: why can't we continue along this promising path forever? One of the big bottlenecks constraining this route is given by the intrinsic complexity of organic materials. The space of computationally feasible materials strongly depends on the level of approximation {used in the simulation}. For example, density functional theory (DFT) currently counts as the most widely used first-principles approach to calculate materials properties. Although, reports towards linearly scaling DFT approaches exist \cite{guerra1998towards}, the computational costs of conventional DFT codes increase as $\mathcal{O}(N^2\log N)$ to $\mathcal{O}(N^3)$. Regarding all materials containing C and H in the crystallographic open database \cite{gravzulis2012crystallography}, their unit cell sizes can be approximated to follow a log-normal distribution with a mean of 222 atoms in the unit cell \cite{borysov2017organic} (Fig. \ref{Fig:data_availability}b).  

While an organic structure is typically {very complicated}, it also exhibits {recurring} local patterns with similar behavior of local functional properties such as charge-, spin-, and orbital-densities or strain profiles. Machine learning models trained on computationally feasible crystal structures could provide reliable predictions of such properties in materials outside the reach of ab initio methods. Such {recurring} patterns also play a central role in machine learning descriptors -- numerical representations of the structures -- for polymers, where the material is built of large molecules composed of many repeating subunits \cite{Huan2015}. {The local environment} around an atom in real space {is} most commonly used to predict material properties, while non-local representations are used to include long-range physics \cite{Grisafi_2019}.

The structure-property relationship of organic molecules was first captured successfully by using handcrafted descriptors such as the Coulomb matrix in 2011 \cite{CoulombMatrix}. This study initiated many works introducing new descriptors that possess desired properties, such as rotational symmetry or uniqueness of the descriptor. Recently, there are two updates to this scheme. Firstly, using more sophisticated techniques from data science, the construction of explicit descriptors is usually no longer a separate step, but rather integrated into the regression model itself. Two examples are SchNet \cite{schutt2017schnet}, a continuous-filter convolutional neural network, and SOAP \cite{soap}, a similarity kernel based on the integration of atomic positions. Secondly, the next step beyond scalar properties is the prediction of higher dimensional objects, such as the density of states, band structure, or charge and magnetization density. This direction is currently actively worked on and progress is promising, for example with respect to reliable charge density predictions \cite{tsubakiquantum, gong}.

Supervised machine learning algorithms approximate an unknown or computationally expensive map between input and output vectors. As such they provide predictive power within the range of the given data. However, attempting extrapolation usually fails. For certain machine learning methods, {for example} Bayesian inference, a large uncertainty in the prediction can provide a sign of extrapolation. Informally speaking, {the} failure of extrapolation in neural networks {can} be understood by the observation that deep nets are essentially performing polynomial regression with high-degree polynomials. The flexibility of capturing any oscillations {within the dataset} provides predictive power, but diverges outside the known data points. However, by training a model on reoccurring local environments and their properties in small organic crystals, an application to complex materials does not necessarily fall into this class. In fact, a sufficiently large dataset containing simple organic molecular crystals will provide the majority of local atomic neighborhoods present in a dataset of complex organic crystal structures. 
\begin{figure*}
    \centering
    \includegraphics[width=0.95\textwidth]{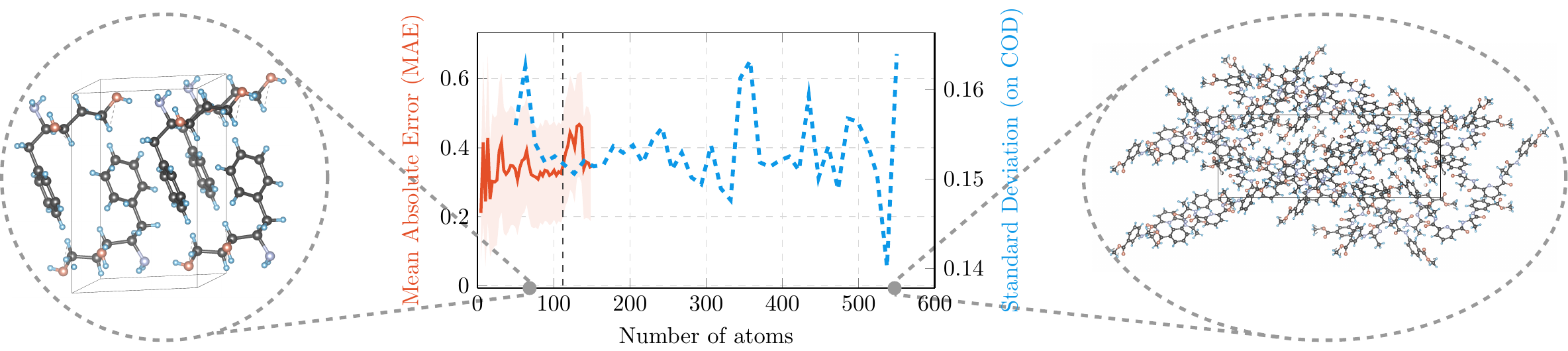}
    \caption{Machine learning techniques are designed for data regression and typically fail when it comes to extrapolation. However, certain machine learning frameworks trained on local environments in materials with feasible unit cell sizes can be applied to predict properties of complex materials having similar local properties. {Using} the example of the band gap we show the $N_{\text{atoms}}$-resolved mean absolute error for an ensemble of machine learning models inside and outside the range contained in the training set. As the model is based on local atomic environments an `extrapolation' towards more complex materials remains fairly stable. }
    \label{Fig:extrapolation}
\end{figure*}

We show this reasoning on the example of the band gap in non-magnetic organic crystals, where we follow the approach of Ref. \cite{olsthoorn2019band}. We train $100$ SchNet models (for hyperparameters refer to \cite{olsthoorn2019band}) on identical training, validation and test data. However, each training of the model starts by a random initiation and therefore ends up with slightly different learnable parameters. We can benefit from this behavior by discussing a collection of trained models, and taking the mean and standard deviation of the band gap prediction. The training and validation dataset only contains materials with fewer than 112 sites, whereas the test set only contains larger materials. The SchNet infrastructure predicts a site contribution based on each local site environment in the unit cell before a final pooling layer gives the scalar result. For intensive properties, such as band gap in this work, an average is used. Our model trained on 25,313 materials taken from the Organic Materials Database achieves a mean absolute error of {approximately $0.4\,\mathrm{eV}$}, which is a reasonable accuracy, as organic crystals are typically large band gap insulators with an average gap of {around 3 eV} \cite{borysov2017organic}. In Fig. \ref{Fig:extrapolation} we show the evolution of the mean absolute error and the standard deviation of the mean absolute error achieved by our ensemble of machine learning models resolved for the number of atoms in the unit cell. While the model is trained on materials with less than 110 materials in the unit cell, we can see that the general trend of our predictive power is also maintained in region of up to 150 atoms in the unit cell. We also evaluate all 100 models on 213,681 materials in the {Crystallographic Open Database}, and show the standard deviation in Fig.~\ref{Fig:extrapolation} remains fairly constant up to about 500 atoms in the unit cell. If we were in a pure extrapolation regime, the predictions of neural networks typically grow wild and as such the standard deviation would increase rapidly. This result shows a promising direction towards predicting materials properties outside the range of materials that can be described with conventional ab initio methods.  

Unsupervised ML allows {identification of the} inner structure in data, {for example} by dimensionality reduction and clustering. It also provides the prospect of generative modeling, providing algorithms which generate authentic fake data following a given statistical distribution. Prominent algorithms are variational autoencoders \cite{kingma2013auto} or Generative Adversarial Networks  \cite{goodfellow2014generative}. Recently, such models {have been applied more frequently to complex} materials and drug design \cite{sanchez2018inverse,jackson2019recent}. This development is a huge improvement towards the hard task of computationally determining an unknown stable molecular or crystal structure. Closing the loop by combining the prediction of stable structures with the prediction of desired properties represents a huge advancement to materials design, significantly extending the available search space.

In summary, we see an increasing interest in highly complex, but fascinating organic crystals. This interest is fostered by an increasing computational power along with the adoption of new tools {that have made it possible to conduct}  sophisticated materials simulations {at the same time as} synthesis and experiment. 
The success of the current approaches {for using} materials informatics largely rests on the established paradigm of scaled up ab-initio methods: {\em structure $\rightarrow$ properties $\leftrightarrow$ database $\rightarrow$ end user and applied machine learning tools. } Perhaps, with the new tools {we have described that use} ensembles of various machine learning models for various materials properties we {will} see the emergence of a new {phase of} computational complex materials research. As such, the outcome of machine learning algorithms could foster the generation of a second generation of databases involving a much larger scale of feasible materials. Such a workflow would be of {a different} form: {\em (artificial intelligence $\rightarrow$) structure $\rightarrow$ artificial intelligence $\rightarrow$ properties $\rightarrow$ database $\rightarrow$ end user.} {This `artificial knowledge' has} the potential to predict materials properties beyond the known constraints of conventional ab initio tools. It would, for example, build the bridge between materials informatics and organized functional biomolecules \cite{simon2019supercharging}. While such a methodology needs to be developed and evaluated in more detail, we believe that this route is a promising candidate towards exascale materials simulations.

\section*{Acknowledgments}
We are grateful for collaboration and discussions with J. Hellsvik, S. S. Borysov. We acknowledge funding from the ERC synergy grant "HERO" (No 810451), University of Connecticut,  VILLUM FONDEN via the Centre of Excellence for Dirac Materials (Grant No. 11744), the Knut and Alice Wallenberg foundation ( 2019.0068) as well as the Vetenskapsrådet (no. 2017.03997). We acknowledge computational resources from the Swedish National Infrastructure for Computing (SNIC) at the the Centre for High Performance Computing (PDC), the High Performance Computing Centre North (HPC2N), and the Uppsala Multidisciplinary Centre for Advanced Computational Science (UPPMAX).

\end{document}